\documentclass{aa}
\usepackage{graphicx}

\begin{document}

\thesaurus{07        
          (07.13.1;  
           02.16.1;  
           02.19.1)} 

\title{A space charge model for electrophonic bursters}

\author{Martin Beech\inst{1} \and Luigi Foschini\inst{2}}
 
 
\institute{Campion College, and Department of Physics, University of Regina, Regina, SK, S4S 0A2,  
Canada (email: Martin.Beech@uregina.ca).
\and
CNR -- ISAO, Via Gobetti 101, I-40129 Bologna, Italy (email: L.Foschini@isao.bo.cnr.it).}
 
\date{Received 9 March 1999; Accepted 7 April 1999}

\authorrunning{M. Beech \& L. Foschini}

\maketitle

\begin{abstract}
The sounds accompanying electrophonic burster meteors are characteristically described
as being akin to short duration ``pops'' and staccato--like ``clicks''. As a phenomenon 
distinct from the enduring electrophonic sounds that occasionally accompany the passage 
and ablation of large meteoroids in the Earth's lower atmosphere, the bursters have 
proved stubbornly difficult to explain. A straightforward calculation demonstrates that in
contradistinction to the enduring electrophonic sounds, the electrophonic bursters
are not generated as a consequence of interactions between the meteoroid ablation plasma 
and the Earth's geomagnetic field. Here we present a novel and hitherto unrecorded
model for the generation of short--duration pulses in an observer's local
electrostatic field. Our model is developed according to the generation of a strong electric 
field across a shock wave propagating in a plasma. In this sense, the electrophonic bursters are
associated with the catastrophic disruption of large meteoroids in the Earth's atmosphere.
We develop an equation for the description of the electric field strength in terms of the 
electron temperature and the electron volume density. Also, by linking the electron line density 
to a meteor's absolute visual magnitude, we obtain a lower limit to the visual magnitude 
of electrophonic burster meteors of $M_{\mathrm{v}}\approx -6.6$, in good agreement with the available 
observations.

\end{abstract}

\section{Introduction}
While Electrophonic meteor sounds have been widely reported throughout recorded
history, they are, none--the--less, a poorly observed phenomena (Keay \& Ceplecha, \cite{KEAY3}). 
By this we mean that the accounts of electrophonic sounds are mostly anecdotal and secondary. 
To our knowledge only two electrophonic meteors have ever been recorded instrumentally. These are the 
fireball events of 1981, August 13th as reported by T. Watanabe and co--authors in Japan 
(see Keay, \cite{KEAY5} for details) and 1993, August 11th as reported by Beech et al. 
(\cite{BEECH1}).

Expressed in terms of two broadly divided classes, electrophonic sounds are either of the 
short duration, or burst type in which a sharp ``click'' of ``pop'' is reported, or of the 
sustained type in which a temporally extended ``rushing'' or ``crackling'' sound is heard 
(Keay, \cite{KEAY4}). For brevity and phenomenological reasons we shall call the short duration 
electrophonic sounds ``bursters''. The essential characteristics of the electrophonic bursters 
are their short durations, $\tau\approx 1$~s, and their piquant impression on the human 
auditory system. 

It is not presently possible to draw any clear statistical inferences from the available data 
on electrophonic sounds. This is due primarily to the fact that it is the local transduction 
conditions that dictate whether or not an electrophonic sound will be heard (Keay, \cite{KEAY1}, 
\cite{KEAY5}). A fireball that some observers report as being electrophonic may be ``silent'' to 
other near--by witnesses simply because the environmental conditions have changed. 
Also, personal ``in--field'' experience has revealed that unsuspecting public observers often 
fail, at least initially, to mention that they heard an associated sound when describing a fireball 
event, thinking that the sounds were either an illusory or irrelevant coincidences. All this 
being said, the literature survey conducted by Kaznev (\cite{KAZNEV}) revealed that from a total of 
888 electrophonic meteor events some 76 (8.5~\%) would qualify for membership in our 
burster category. The survey by Keay (\cite{KEAY4}) indicates that 31 (10~\%) out of the 301 
events considered would qualify as bursters. We note also that both of the instrumentally 
observed electrophonic meteors fall into the short duration burster class. The observations
also indicate that electrophonic burster meteors must be very bright. Indeed, the 1981, August 13th 
electrophonic fireball event recorded in Japan had an estimated visual magnitude of $-6$, while
the fireball of the 1993, August 11th had an estimated visual magnitude of $-10$. 

Keay (\cite{KEAY1}) and Bronshten (\cite{BRONSHTEN}) have developed a robust theory to explain the 
extended ``rushing'' or ``crackling'' electrophonic sounds. The key physical mechanism 
identified in the production of these sounds is the freezing--in and ``twisting'' of the 
geomagnetic field in the turbulent wake behind a large meteoroid. In this mechanism it is 
the release of the strain energy in the geomagnetic field that produces very low frequency 
(VLF) radio waves and these, depending upon the local environmental conditions, are 
transducted into audible sounds. The generation of a VLF radio wave signal will proceed 
provided that the Reynolds number in the meteor ablation column is greater than $10^{6}$ (i.e., 
the ablation column is turbulent) and that the magnetic Reynolds number is 
concomitantly greater than unity. Irrespective of the environmental surroundings, if the 
Reynolds number conditions are satisfied, then an appropriately designed receiver should 
detect the VLF radio signal. These conditions can be useful in order to evaluate the dimension of the 
meteoroid (Beech, \cite{BEECH}).

The mechanisms responsible for producing electrophonic bursters have not been as 
straightforward to annotate as those for the extended sounds. However, the inherent 
characteristics of burster events suggest that they relate to catastrophic rather than 
on--going events in the atmospheric ablation of a meteoroid. This phenomenological 
argument suggests an association between electrophonic bursters and meteor flares and 
terminal detonations. 

\section{The magnetic cavity model}
A compelling ``first guess'' model for the generation of an electrophonic burster is the 
excavation of a cavity in the Earth's magnetic field. This could be achieved by the 
propagation of a highly ionized blast wave into the static geomagnetic field, as shown by Karzas \& Latter
(\cite{KARZAS}) for nuclear airbursts. The energy density of the blast wave produced by the 
catastrophic disruption of an ablating meteoroid is typically much lower than that of a nuclear detonation.
Indeed, as we show below, the energy required to excavate a magnetic cavity capable of producing an
electrophonic burster is characteristic of that of an impacting asteroid, rather than a detonating meteoroid.
To first order, the power radiated will be given by:

\begin{equation}
P=U_{\mathrm{m}}\frac{4\pi}{3}R^{3} 
\label{e:pow}
\end{equation}

\noindent where $U_{\mathrm{m}}$ is the geomagnetic field energy density 
($U_{\mathrm{m}} = B^{2}/2\mu_{0}\approx 10^{-3}$~J/m$^{3}$) and $R$ is the radius of the cavity. 
The cavity radius is set according to the distance at which the conductivity drops below the 
level for magnetic field entrapment. Keay (\cite{KEAY1}) has argued that human electrophonic 
hearing begins once the electrostatic field strength variations exceed 160~V/m (peak--to--peak). 
For a fireball to produce such a variation in the electrostatic field, at 
a distance of say 40~km, a power output of some $2\times10^{11}$~W is required
(assuming a dipolar radiation field). In order to produce the required amount of power on the 
typical burster time scale, the detonating meteoroid would have to excavate 
a cavity of radius 37~km in about one second. In order to produce such a large cavity the 
detonating meteoroid would have to deposit a very large amount of energy into the expanding blast
wave. Indeed, as we show below, an unreasonably large amount of energy is 
required. From Taylor (\cite{TAYLOR1}, \cite{TAYLOR2}) we find that the propagation speed $V$ [km/s]  
of a blast wave will be of order

\begin{equation}
V = 4.13\times 10^{-6}\bigl(\frac{E}{\rho t^{3}}\bigr)^{1/5}
\label{e:speed}
\end{equation}

\noindent where $\rho$ is the atmospheric density at the detonation height [kg/m$^{3}$] and $E$ is the energy 
deposited by the detonating meteoroid [J]. Assuming a detonation height of 30~km, 
we have $\rho\approx 0.02$~kg/m$^{3}$ and $E\approx 10^{33}$~J. Clearly, the required energy 
deposition in the magnetic cavity model is unrealistically high. Indeed, it is characteristic of that 
expected from a large impacting asteroid rather than a large meteoroid.
In this respect, we need to look for other burster generation mechanisms. In the section below we 
present a novel model for the production of electrophonic bursters, building our arguments upon 
the fact that meteoroids enter the Earth's atmosphere at hypersonic velocities.

\section{Space charge separation by shock waves}
When a meteoroid enters the Earth's atmosphere it moves at hypersonic 
speeds, that is with Mach number greater than 5. Hence, behind the 
bow shock the effect of ionization becomes very important (for a 
brief description of hypersonic flow around a meteoroid see Foschini,  
\cite{FOSCHINI1} and references therein). Other effects, such as 
ablation, contribute to enhance the presence of charged particles in 
the fluid around the meteoroid (for a review see Ceplecha et al., 
\cite{CEPLECHA}). Moreover, the presence in meteoroids of alkaline and 
alkaline--earth metals, which easily ionize, results in the rapid formation 
of a plasma sheet around the meteoroid body (see Foschini, \cite{FOSCHINI2}).

When the meteoroid is large enough to create an energetic airburst, the 
shock wave propagates in the plasma.  Owing to the 
presence of large gradients of pressure, temperature and other 
quantities, across the shock, and taking into account the very 
different masses of electrons and ions, there is a strong diffusion of 
the electron gas with respect to the ion gas.  However, the diffusion in a 
plasma is quite different from the diffusion in a neutral gas, because 
a small change in the charge neutrality gives rise to a strong 
electric field, which in turn tends to restore the neutrality and to 
prevent further diffusion.  

In order to estimate the order of magnitude of the electric field 
generated by a shock wave we refer to the book 
by Zel'dovich \& Raizer (\cite{ZELDOVICH}).  For the sake of the 
simplicity we assume that the ions are singly ionized and, therefore, 
$n_{\mathrm{e}}=n_{\mathrm{i}}=n$, where $n_{\mathrm{e}}$ and $n_{\mathrm{i}}$ are the electron and ion volume 
density respectively. 
Let $x$ be the dimension of the compression shock, that 
is the characteristic length along which macroscopic variables have 
strong changes.  In this region, the shock wave generates a local 
difference $\delta n = n_{\mathrm{i}}-n_{\mathrm{e}}$ between the ion and electron 
densities.  The electric field produced by the space charge $e\cdot \delta 
n$, where $e$ is the elementary electric charge, can be calculated from  
Gauss' law (we are interested in modulus only):

\begin{equation}
	E=\frac{e\cdot\delta n \cdot x}{\epsilon_{0}}
	\label{e:gauss}
\end{equation}

\noindent and the potential difference across the shock will be 
$\delta \phi = Ex$. In the absence of other external fields (we can 
neglect the Earth's magnetic field) the separation of ions and electrons 
is maintained by thermal motion only. Therefore the electron 
potential cannot exceed $kT$, where $k$ is Boltzmann's constant 
and $T$ is the temperature. Therefore:

\begin{equation}
	\delta \phi \approx kT \rightarrow \frac{\delta n}{n}\approx 
	\frac{\epsilon_{0}kT}{e^{2}nx^{2}}=\frac{\lambda_{\mathrm{D}}^{2}}{x^{2}}
	\label{e:dn}
\end{equation}

\noindent where $\lambda_{\mathrm{D}}$ is the Debye length, that establishes the 
characteristic dimension where the electrostatic force dominates over the 
thermal force. Eq.~(\ref{e:dn}) shows that a strong separation of 
charges, that is $\delta n/n \approx 1$, occurs when the 
characteristic dimension of the shock is of the order of the Debye 
length.

The largest gradients in a plasma appear in the viscous compression 
shock, where the macroscopic variables undergo a large change
on a scale length of the order of the mean free path of the charged particles:

\begin{equation}
	x=l=\frac{v}{\nu_{\mathrm{ei}}}
	\label{e:mfp}
\end{equation}

\noindent where $v$ is the electron mean speed and $\nu_{\mathrm{ei}}$ is the 
electron--ion collision frequency. In a plasma, the Maxwellian 
distribution speed is established quite quickly, even though 
there is a rather slow energy exchange between the electrons and ions. 
The effect is that there are two gases (electron gas and ion gas) with different 
temperatures, but both with Maxwellian distributions. We then 
consider the electron temperature as reference. The mean Maxwellian 
speed is:

\begin{equation}
	v=\sqrt{\frac{8kT}{\pi m_{\mathrm{e}}}}
	\label{e:maxw}
\end{equation}

\noindent where $m_{\mathrm{e}}$ is the electron rest mass. 

Concerning the collision frequency, we can note that interactions between 
ions and electrons are more frequent than other interactions, owing to the 
electrostatic field. Particularly for high electron densities, such 
as during airbursts, $\nu_{\mathrm{ei}}$ dominates over all other frequencies (Foschini,  
\cite{FOSCHINI2}), so that we can consider the plasma as fully ionized. 
We can therefore use the following formula, valid 
for a singly ionized plasma (Mitchner \& Kruger, \cite{MITCHNER}):

\begin{equation}
	\nu_{\mathrm{ei}}=n\frac{4\sqrt{2\pi}}{3}\sqrt{\left(\frac{m_{\mathrm{e}}}{kT}\right)^{3}}
	\left(\frac{e^{2}}{4\pi \epsilon_{0}m_{\mathrm{e}}}\right)^{2}\ln \Lambda
	\label{e:freq}
\end{equation}

\noindent where $\ln \Lambda$ is the Coulomb logarithm. This equation 
is valid when $\ln \Lambda >> 1$, that is under the condition that the gas is a plasma.

We can now obtain an expression for the electric field generated 
across a shock wave in a plasma. We have to substitute 
Eq.~(\ref{e:maxw}) and Eq.~(\ref{e:freq}) in Eq.~(\ref{e:mfp}). We then have:

\begin{equation}
E=\frac{\delta \phi}{l}\approx \frac{kT}{el}=\frac{ne^{3}}{24kT\pi 
\epsilon_{0}^{2}}\ln \Lambda 
\label{e:efield}
\end{equation}

\begin{figure}[ht]
\centering
\includegraphics[angle=270, width=\linewidth]{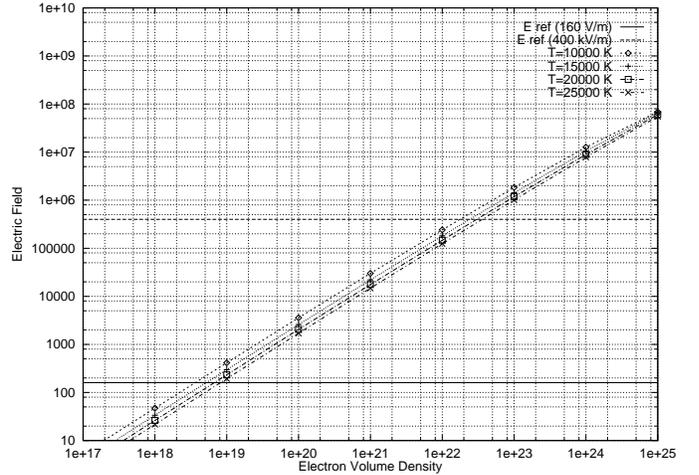} 
\caption{The electric field generated by space charge separation in shock waves. The electric field value is in [V/m], while the
electron volume density is in [m$^{-3}$]. Reference electric field of
160~V/m and 400~kV/m are also indicated (see text for details). 
Lines are plotted for several values of the temperature.}
\label{electric}
\end{figure}

Fig.~\ref{electric} shows the electric field, calculated according to Eq.~(\ref{e:efield}), versus the
electron volume density for various temperatures. The reference electric field $E=160$~V/m is also shown.
We can note that we obtain the required field with $n\approx 4\times 10^{18}$~m$^{-3}$ and
$T=10^{4}$~K.

\section{Discussion}
For an evaluation of the absolute visual magnitude corresponding to a given electron volume density, we
can use the formula cited by Allen (\cite{ALLEN}), valid for very bright bolides, and adapted for our 
purposes:

\begin{equation}
M_{\mathrm{v}}=35.5 - 2.5\log\alpha_{\mathrm{z}} - \delta M 
\label{e:vismag1}
\end{equation}

\noindent where $\alpha_{\mathrm{z}}$ is the electron line density [cm$^{-1}$] corrected for the zenith distance
and $\delta M$ is a correction factor depending on meteoroid speed. It is worth noting that meteoroid
producing bolides are almost all of asteroidal origin (Jopek et al., \cite{BOLIDES}, Foschini et al., \cite{FOSCHINI3}), 
therefore we can consider $\delta M = 1.9$, that is the correction for a speed of $20$~km/s. Moreover, we have to 
consider volume density instead of line density, therefore we have to insert a correction factor for $\alpha$. 
Since we are only interested in the peak magnitude we consider the greatest electron volume density only,  
namely, the density appropriate to a circular cylinder with a radius equal to the initial
train radius of about 1~m.
We can neglect the zenith correction, owing to the fact that we are dealing with very bright bolides.
Then, we obtain:

\begin{equation}
M_{\mathrm{v}}=35.5 - 2.5\log(3.2\times 10^{-3}\cdot n) - 1.9
\label{e:vismag2}
\end{equation}

\begin{figure}[ht]
\centering
\includegraphics[angle=270, width=\linewidth]{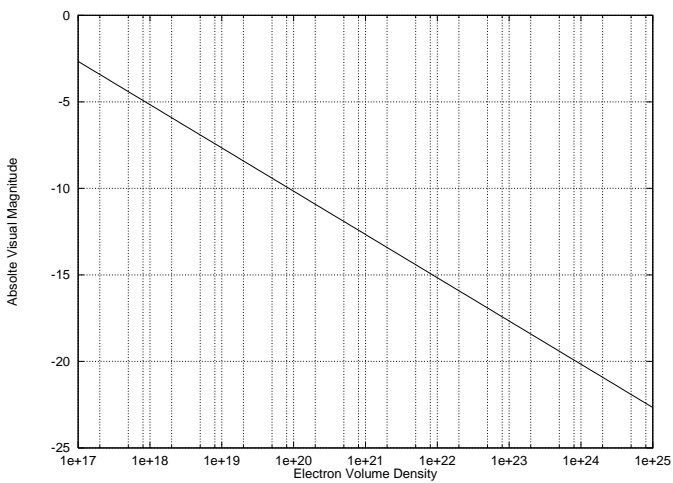} 
\caption{Absolute visual magnitude as a function of electron volume density. The
electron volume density value is in [m$^{-3}$].}
\label{magn}
\end{figure}

Now, we can calculate the minimum visual magnitude corresponding to the minimum electron volume density
necessary to produce the required electric field. From Fig.~\ref{electric} we have $n\approx 4\times
10^{18}$~m$^{-3}$. Substituting this value in Eq.~(\ref{e:vismag2}) we obtain $M_{\mathrm{v}}\approx -6.6$. For the sake of
simplicity we show in Fig.~\ref{magn} the plot of $M_{\mathrm{v}}$ as a function of $n$, calculated according to 
Eq.~(\ref{e:vismag2}). We specifically note that at an electron volume density of $10^{20}$~m$^{-3}$, an absolute visual 
magnitude of $-10$ is implied (see Fig.~\ref{magn}) and an electric field strength of about $2500$~V/m is 
expected (see Fig.~\ref{electric}). These numbers are, in fact, in excellent agreement 
with the measurements collected during the August 11th, 1993 fireball event, 
where the electric field strength was calculated to be greater than $2000$~V/m (Beech et al., \cite{BEECH1}).

The reference electric field of 160~V/m was obtained by Keay (\cite{KEAY1}) by means of experiments with human
beings. Later on, Keay \& Ostwald (\cite{KEAY6}) made some experiments in order to measure the acoustic response 
of several objects and materials by means of an applied electric field of 400~kV/m. If we consider this value 
as reference, the implied fireball magnitude is $-16$.

\section{Conclusions}
In this Letter we have presented a novel and hitherto unreported model for the 
generation of short--duration electrophonic bursters. The model assumes  
the catastrophic disruption of a large meteoroid and the 
subsequent separation of electrons and ions by an energetic shock wave. Since 
meteoroids enter the Earth's atmosphere with hypersonic velocities, an airburst 
detonation results in the formation of a shock wave that propagates in the 
plasma boundary. As a result of the large temperature and pressure gradients 
across the shock, there is significant diffusion of the electron gas with 
respect to the ion gas, with the result that an electric field is produced 
by the space charge separation. It is the rapid variation in the electric 
field strength that results in the potential generation of electrophonic sounds.

At present the observations only afford one case (the August 11th, 1993 
fireball event) in which actual measurements can be compared against the 
predictions. We are pleased to find, however, that in this one case there is 
an excellent agreement between the predicted electric field strength variation 
(as described by Eq.~\ref{e:efield}) and the measurements.

While our model was specifically developed to explain the electrophonic 
burster events, we note that the same basic mechanism may also operate with 
respect to producing extended electrophonic sounds. That is, if shock waves are 
produced within the hypersonic flow around a large ablating meteoroid, 
the space charge separation mechanism can ``run'' in a temporally extended 
fashion. We hope to investigate this situation in future work.

\begin{acknowledgements}
We wish to thank the referee, C.~Keay for useful comments. A thank also to 
P.~Farinella for constructive review.
\end{acknowledgements}

\end{document}